**Title: Prediction of soil fertility parameters using USB-microscope imagery and portable X-ray fluorescence spectrometry**

**Authors:** Shubhadip Dasgupta[a,b], Satwik Pate[a], Divya Rathore[a], L.G. Divyanth[c], Ayan Das[d], Anshuman Nayak[a], Subhadip Dey[a], Asim Biswas[e], David C. Weindorf[f,*], Bin Li[g], Sérgio Henrique Godinho Silva[h], Bruno Teixeira Ribeiro[h], Sanjay Srivastava[i], Somsubhra Chakraborty[a]

[a] *Agricultural and Food Engineering Department, Indian Institute of Technology Kharagpur, Kharagpur 721302, India*

[b] *Department of Agricultural Chemistry and Soil Science, Bidhan Chandra Krishi Viswavidyalaya, West Bengal 741252, India*

[c] *Center for Precision and Automated Agricultural Systems, Department of Biological Systems Engineering, Washington State University, Prosser, WA 99350, USA*

[d] *Space Applications Centre, Indian Space Research Organisation (ISRO), Ahmedabad 380015, India*

[e] *School of Environmental Sciences, University of Guelph, 50 Stone Road East, Guelph, Ontario, N1G 2W1, Canada*

[f] *School of Earth, Environment, and Sustainability, Georgia Southern University, Statesboro/Savannah, GA, USA*

[g] *Department of Experimental Statistics, Louisiana State University, Baton Rouge, LA 70802, USA*

[h] *Department of Soil Science, Federal University of Lavras, Lavras, Minas Gerais, Brazil*

[i] *Indian Council of Agricultural Research (ICAR)-Indian Institute of Soil Science, Bhopal, Madhya Pradesh 462038, India*

**Corresponding Author:**

David C Weindorf, Ph.D.
Professor and Vice President for Research and Economic Development
Georgia Southern University, Statesboro/Savannah, GA, USA
Email: dweindorf@georgiasouthern.edu

## ABSTRACT

This study investigated the use of portable X-ray fluorescence (PXRF) spectrometry and soil image analysis for rapid soil fertility assessment, with a focus on key indicators such as available boron (B), organic carbon (OC), available manganese (Mn), available sulfur (S), and the sulfur availability index (SAI). A total of 1,133 soil samples from diverse agro-climatic zones in Eastern India were analyzed. The research integrated color and texture features from microscopic soil images, PXRF data, and auxiliary soil variables (AVs) using a Random Forest model. Results showed that combining image features (IFs) with AVs significantly improved prediction accuracy for available B (R2 = 0.80) and OC (R2 = 0.88). A data fusion approach, incorporating IFs, AVs, and PXRF data, further enhanced predictions for available Mn and SAI, with R2 values of 0.72 and 0.70, respectively. The study highlights the potential of integrating these technologies to offer rapid, cost-effective soil testing methods, paving the way for more advanced predictive models and a deeper understanding of soil fertility. Future work should explore the application of deep learning models on a larger dataset, incorporating soils from a wider range of agro-climatic zones under field conditions.

**Keywords:** Microscope; Organic carbon; Soil nutrients; Image processing; PXRF; Random forest

## 1. INTRODUCTION

Understanding the variability of soil properties is paramount for enhancing agricultural productivity and sustainability. In the context of India, particular attention to specific micronutrients and organic carbon (OC) is crucial due to their significant impact on crop yields and soil health. Notably, available B, organic carbon (OC), available Mn, available S, and sulfur availability index (SAI) are essential for plant growth, affecting various physiological and biochemical processes (Minasny and McBratney, 2016; Moinuddin et al., 2017; Zenda et al., 2021). However, Indian soils are notably deficient in these key micronutrients, with alarming deficiencies reported in Mn, B, and S (Dasgupta et al., 2023; Saha et al., 2019; Singh, 2008; Behera et al., 2009; Majumdar et al., 2014; Shukla et al., 2021). This situation underscores the importance of accurately assessing and managing these soil properties to improve crop productivity. Custom fertilization strategies enhance crop yields and reduce costs, promoting soil health. However, the scarcity of soil testing facilities, particularly in developing nations, hinders precise nutrient management. As a result, there is a need for affordable soil fertility assessment tools (Dasgupta et al., 2021; Goswami et al., 2023, Choudhury et al., 2023).

To address the challenges of soil variability and nutrient management, three distinct methods have emerged as promising alternative tools for predicting soil properties: traditional soil color assessment, digital image analysis, and portable X-ray fluorescence (PXRF) spectrometry. Traditional soil color assessment, using the Munsell soil color chart (MSCC), has long been a qualitative indicator of pedogenic and chemical processes within the soil (Rowe, 2005). However, the subjective nature of this method limits its precision and repeatability (Mouazen et al., 2007; Swetha et al., 2022). Conversely, digital image analysis offers a quantitative approach, leveraging smartphone-captured images analyzed through machine learning (ML) and deep learning (DL) algorithms. This method enables the extraction of features such as color, texture, and morphology, offering a more efficient and cost-effective means of soil testing (Aitkenhead et al., 2016a; Fu et al., 2020; Gorthi et al., 2021; Qi et al., 2019; Swetha et al., 2020; Taneja et al., 2021; Viscarra Rossel et al., 2008). PXRF spectrometry adds another layer of analysis, providing rapid, accurate readings of elemental composition, nutrient levels, and heavy metal contamination, beneficial for in-field applications (Chakraborty et al., 2016; Dasgupta et al., 2022). Additionally, the integration of computer vision-based image analysis with a universal serial bass (USB) microscopy offers a microscopic perspective that has proven beneficial for interpreting soil features (Sudarsan et al., 2018; Sudarsan et al., 2016). Yet, the application of a USB microscope as a proximal soil sensor for assessing soil fertility parameters needs further exploration.

While each method offers distinct advantages, they also come with limitations. Subjective assessment of soil color can result in inconsistencies, digital image analysis requires specific conditions for image capture, and PXRF, though precise, is unable to measure soil OC and available nutrient fractions directly. Recognizing these challenges, there is a growing interest in the fusion of these technologies to enhance soil property prediction. Integrating outputs from multiple sensing technologies, known as sensor fusion, has shown potential in improving predictions by combining data from different sensors (Aldabaa et al., 2015; Duda et al., 2017; Swetha and Chakraborty, 2021; Wang et al., 2015). However, the accuracy and reliability of this integrated approach across diverse soil types and environmental conditions warrant further exploration (Andrade et al., 2022).

Given these considerations, our study aims to predict available B, OC, available Mn, available S, and SAI across five Indian states and in the Indo-Gangetic Plain (IGP) using various approaches, either alone or in combination of color and image texture features from USB microscope-based soil images (IFs), auxiliary variables (AVs) such as soil agro-climatic and physiographic information, and PXRF elemental data. We hypothesize that combining USB microscope-based soil image and PXRF data with AVs enhances soil fertility predictions compared to using USB microscope images or PXRF data in isolation. Through this work, we aim to contribute to the advancement of precision agriculture and sustainable soil management practices.

## 2. MATERIALS AND METHODS

### *2.1. Study area*

Conducted in the IGP, spanning five Indian states and characterized by rivers forming a vast floodplain (Shukla et al., 2021), the study area covers over 45,000 km$^2$. Divided into six Agro-Climatic Zones (ACZs) across nine districts—Gangetic Alluvial Zone (GAZ), Coastal Saline Zone (CSZ), Vindhyachal Alluvial Zone (VAZ), Terai-Teesta Alluvial Zone (TAZ), Northern Hilly Zone (NHZ), and Red and Laterite Zone (RLZ)—the region varies in elevation (-1 to 980 meters above sea level) and climate (Fig. 1). The geographical coordinates range from 24° 48' 27" N to 21° 41' N latitude and 86°06' 50’’E to 89° 25' 38’’E longitude, encompassing distinct Köppen climatic classes (Aw, Cwa, and Cwb). The CSZ, marked by periodic saline tidal water intrusion through rivers (Sarkar et al., 2022), covers half of West Bengal state. Known as India's rice-bowl, West Bengal faces challenges with a rice-based mono-cropping system and limited micronutrient fertilizer use, leading to deficiencies. Besides rice, the region cultivates winter vegetables, pulses, potatoes, and groundnuts (Dasgupta et al., 2022).

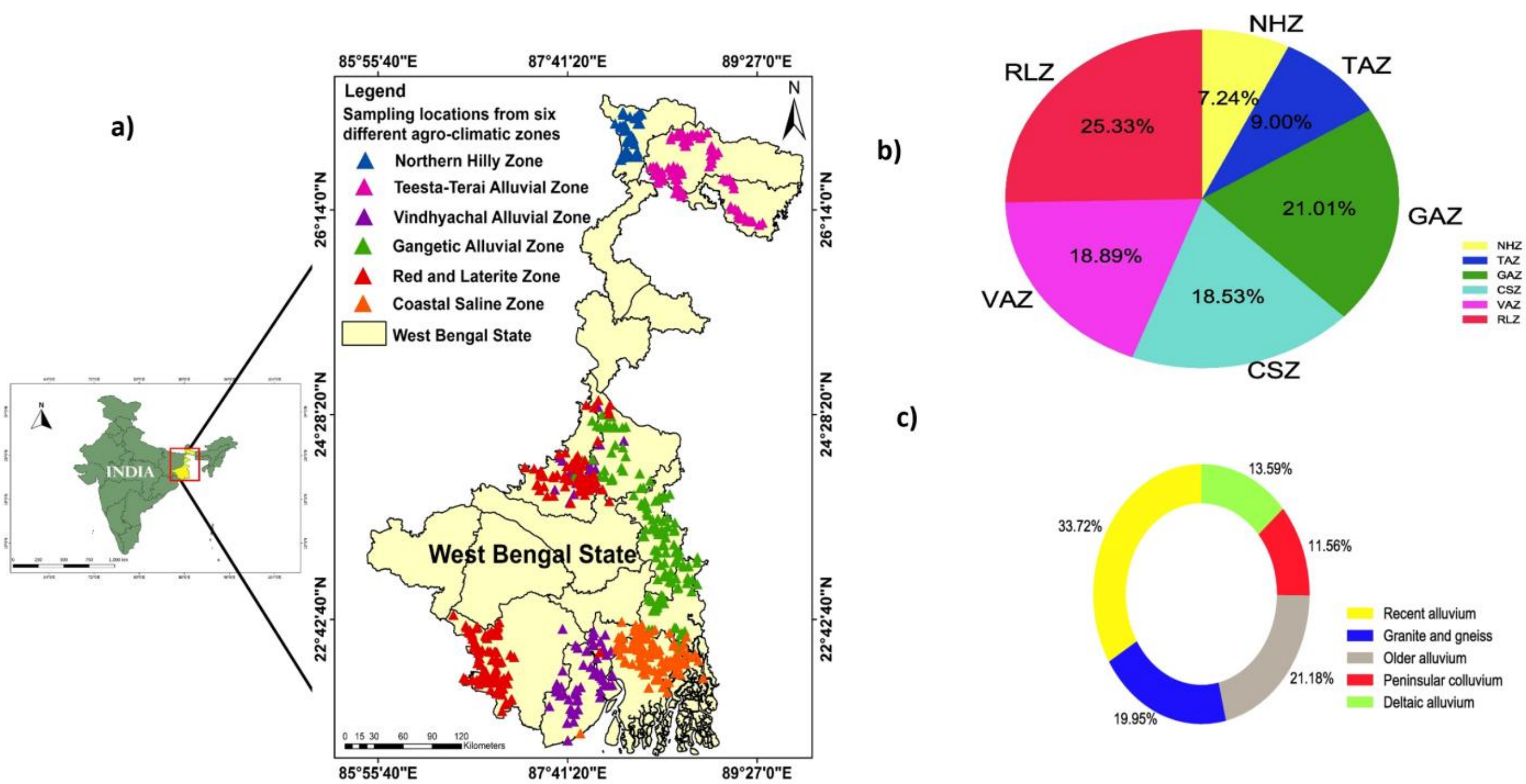

**Figure 1**: a) Map of the study region in the West Bengal state, India b) % distribution of soil samples collected from the six agro-climatic zones of the study region. c) % distribution of soil samples collected from the five soil parent materials of the study region.

### *2.2. Soil sampling and laboratory analysis*

Between 2017 and 2018, about 1,133 surface soil samples (0-30cm depth) were collected from rice fields in November to December, covering rainfed and irrigated agricultural lands of various sizes (Table 1). Using stratified random sampling, soils were obtained from 68 distinct soil series, representing the variability across four ACZs with three soil orders and four parent materials. Maintaining consistency, about 20-25 soil samples were collected from each series, spaced approximately 3 km apart. These samples were carefully sealed in plastic bags and geolocated with a Garmin handheld GPS (Garmin Ltd., KS, USA) for precision. Categorized by locations—GAZ (n=238), CSZ (n=210), VAZ (n=214), NHZ (n=82), TAZ (n=102), and RLZ (n=287)—the soil library comprised 419 Inceptisols, 269 Entisols, and 445 Alfisols samples. Reflecting three productivity potentials (low, medium, and high), the samples underwent air-drying, grinding, and sieving (2 mm).

**Table 1:** Description of parent-materials, soil order, land use, and productivity potential along with the number of samples collected from six agro-climatic zones of West Bengal, India.

| Agro-Climatic Zone | Parent Material | Soil Order | Land Use | Productivity Potential |
|---|---|---|---|---|
| Northern Hilly | Recent Alluvium:62<br>Granite and Gneiss: 20 | Inceptisols:48<br>Entisols: 34 | Tea and orange plantation, rainfed rice, maize, vegetables. | Medium: 44<br>Low: 38 |
| Teesta-Terai Alluvial | Recent Alluvium: 78<br>Older Alluvium: 24 | Inceptisols: 66<br>Entisols: 36 | Tea plantation and other horticultural crops like orange, pineapple; rice, jute, tobacco, and vegetables. | High: 4<br>Medium: 30<br>Low: 28 |
| Gangetic Alluvial | Recent Alluvium: 150<br>Older Alluvium: 58<br>Peninsular Colluvium: 30 | Inceptisols: 119<br>Entisols: 75<br>Alfisols: 44 | Most of the agricultural and horticultural crops | High: 132<br>Medium: 82<br>Low: 24 |
| Coastal Saline | Deltaic Alluvium: 154<br>Recent Alluvium 38<br>Older Alluvium: 18 | Inceptisols: 77<br>Entisols: 92<br>Alfisols: 41 | Salt resistant rice, oilseed, pulses, and few winter vegetables | High: 36<br>Medium: 118<br>Low: 24 |
| Vindhyachal Alluvial | Recent Alluvium 54<br>Older Alluvium: 60<br>Granite and Gneiss: 35<br>Peninsular Colluvium: 65 | Inceptisols: 71<br>Entisols: 32<br>Alfisols: 111 | Rice, potato, groundnut, pulses, and other winter vegetables | High: 78<br>Medium: 96<br>Low: 40 |
| Red and Laterite | Older Alluvium: 80<br>Granite and Gneiss: 171<br>Peninsular Colluvium: 36 | Inceptisols: 38<br>Alfisols: 249 | Rice, pulses, and fruit crops | High: 47<br>Medium: 72<br>Low:168 |

[a]Nayak et al. (2001)

In the laboratory, OC analysis followed the Walkley and Black (1934) chromic acid digestion method. Available B was determined using hot water and azomethine-H colorimetry following the method described by Berger and Truog (1940). Available S was measured using Calcium chloride ($CaCl_2$) extraction, with turbidity readings taken at 440 nm using a UV-Vis spectrophotometer. Plant-available Mn was extracted with Diethylenetriamine pentaacetate (DTPA) and quantified via atomic absorption spectrophotometer (Lindsay and Norvell, 1978). The SAI, crucial for tropical soils, was calculated according to the method proposed by Donahue et al. (1977) and following protocols outlined by Padhan et al. (2016) and Dasgupta et al. (2022).

### *2.3. Soil image acquisition*

For capturing soil images, air-dried and finely ground soil samples were evenly spread in petri dishes, with approximately 25 g of soil placed in each dish at a thickness of 10 mm. To ensure accurate imaging without shading effects from microtopography, the soil was flattened using a rubber mallet. A compact and cost-effective Koolertron 5MP 20-300X USB-microscope with 300x magnification was utilized, featuring sturdy construction suitable for outdoor use and adjustable LED lights for optimal illumination. Micro Capture Pro software was used for image acquisition, offering controlled illuminations to ensure accurate soil analyses. The LED intensity was set at 75%, preventing overheating while maintaining ample brightness. A custom microscope holder isolated the field of view (FOV) for each soil sample, with a precise FOV of 2 × 1.7 mm. Images were captured at 90° angle to ensure comprehensive coverage, with three replicated shots for each soil sample, rotated 45° each time (Fig. 2). The resulting images had dimensions of 4000 × 3000 pixels and were saved in JPEG format, with an average size of around 5 megabytes.

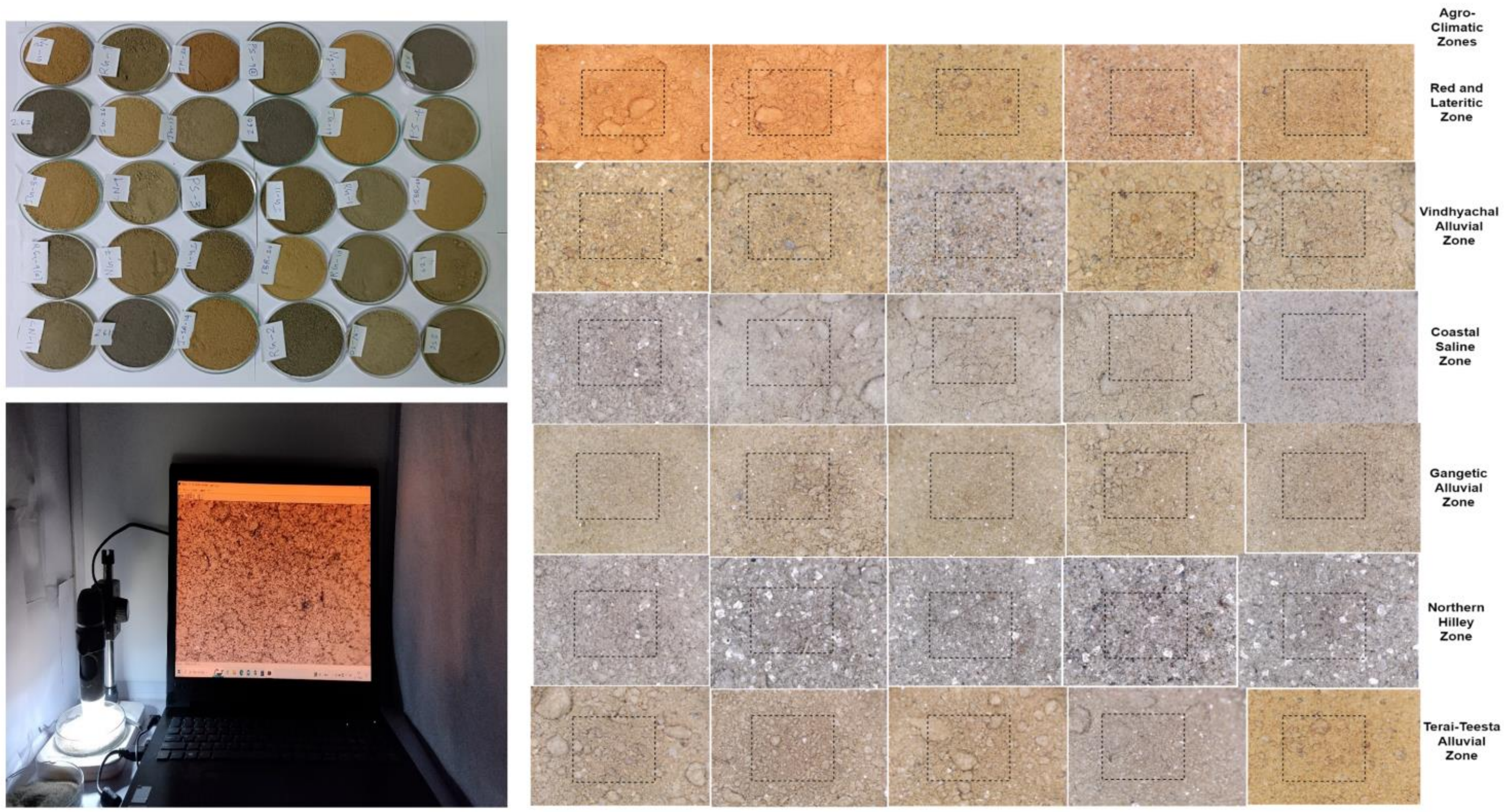

**Figure 2:** Universal Serial Bus (USB) microscope-based soil image acquisition setup showing representative samples from different agro-climatic zones of West Bengal, India.

### *2.4. Scanning via PXRF*

Following the methods outlined by Weindorf and Chakraborty (2020), an Elvatech ProSpector LE PXRF spectrometer was utilized to scan finely ground, air-dried soil samples (<2 mm). The instrument was operated in "Soil Mode" and calibrated with a factory calibration stainless-steel shutter, enabling detection of multi-elemental composition in mg $kg^{-1}$. A total of 19 elements (Ca, K, Fe, Mn, Rb, Zr, Zn, Ti, Ba, Cr, Cu, Pb, Ni, Ag, Sn, V, Sr, Sb, and Ga) were measured in their total elemental content, serving as predictors for targeted soil fertility parameters. To validate the PXRF elemental values, four National Institute of Standards and Technology (NIST) certified reference material (CRM) soil samples underwent scanning. The validation process involved calculating the percentage recovery (% of recovery = 100 × PXRF reported content/certified content) and determining the correction factor following Dasgupta et al. (2022). An element-wise average correction factor (ACF) was computed from the results and applied to the entire set of Compton-normalized PXRF elemental data, following the procedure by Chakraborty et al. (2019). For 15 out of the 19 elements, specific ACFs were utilized, and these were then multiplied with the raw PXRF data for further statistical analysis. For this study, the ACFs used for 15 elements out of 19 from PXRF were as follows: K (1.01/1.54/0.91/1.30/1.19); Ca (0.79/1.25/0.88/1.34/1.06); Fe (0.96/1.26/0.87/1.19/1.07); Mn (0.92/1.26/0.59/1.15/0.98); Rb (0/1.17/0.27/0/0.72); Zn (1.38/1.28/0.97/1.16/1.20); Cu (1.41/1.29/1.01/1.15/1.22); Cr (0.87/1.52/0.36/1.25/1.00); Ti (1.03/1.45/3.25/1.28/1.75); Ni

(0/0.48/0.98/0.63/0.60); Ag (0/0/0.74/0/0.74); Ba (0/0.99/1.16/0.83/0.99); V (0.80/0.29/0.47/0.20/0.44); Ga (0/0/1.51/1.72/1.62); and Pb (1.24/1.16/0.68/0/1.03).

### *2.5. Soil image analysis and feature extraction*

A total of 1,133 soil microscope images were utilized to build the prediction models, with a focus on image texture and color attributes as primary features. All analyses were performed using Matlab (R2022a, Mathworks Inc., Waltham, MA) software, with its statistics and machine learning, image processing, and deep learning toolboxes. Several previous studies have successfully employed image texture features derived from the Gray Level Co-occurrence Matrix (GLCM) and Gray Level Run-Length Matrix (GLRLM) in pattern recognition research (Aygun and Gunes, 2017; Divyanth et al., 2023; Mathew et al., 2022; Nadimi et al., 2023). Herein, textural features were extracted from the GLCM and GLRLM of the individual color channels (R, G, and B) (Fig. 3). The GLCM measures the frequency of pixel value (or gray levels) combinations in grayscale images, providing insights into spatial relationships between pixels. GLRLM, on the other hand, analyzes the lengths and frequencies of consecutive runs of pixels with similar intensities that are aligned in the same direction within the image. Texture feature calculations utilize the information from the GLCM and GLRLM to quantify variations in image texture (pixel values) at the point of interest. Before computing the matrices, the image was quantized into 16 gray levels (Fig. 3), which was determined after experimenting with levels of 8, 16, 32, and 64. After quantization, four GLCMs and four GLRLMs were computed for each color channel, corresponding to the orientations $\Theta_j \in [0°, 45°, 90°, 135°]$. Additionally, four statistics were extracted from the GLCMs and eleven statistics from the GLRLMs, which served as the textural features of the image (see Table 2 for the list of features). The mathematical formulas for these statistical measures are detailed in Divyanth et al. (2023) and can be easily implemented using built-in functions (*graycomatrix*; *graycoprops*) available in MATLAB's Image Processing Toolbox. Color features were derived from the individual channels of the image in HSV and L*a*b* color spaces, along with RGB images, yielding mean, median, and standard deviation values. Hence, in total, 231 IFs were extracted, comprising 204 textural and 27 color features (Table 2).

**Table 2:** Image features (IFs) extracted from the microscope images of soil samples collected from six agro-climatic zones of West Bengal, India.

| **Image feature/statistic measure** | **Color spaces used for feature extraction** | **Total number of features** |
|---|---|---|
| **Textural features from GLCM** | | |
| Correlation<br>Dissimilarity<br>Homogeneity<br>Contrast | RGB | 72 |
| Energy<br>Angular second moment | | |
| **Textural features from GLRM** | | |
| Short-run emphasis (SRE)<br>Long-run emphasis (LRE)<br>Gray level non-uniformity (GLN)<br>Run length non-uniformity (RLN)<br>Run percentage (RP)<br>Low gray level run emphasis (LGRE)<br>High gray level run emphasis (HGRE)<br>Short-run low gray level emphasis (SRLGE)<br>Short-run high gray level emphasis (SRHGE)<br>Long-run low gray level emphasis (LRLGE)<br>Long-run high gray level emphasis (LRHGE) | RGB | 132 |
| **Color features** | | |
| Mean<br>Median<br>Standard deviation (SD) | RGB, HSV, L*a*b* | 27 |

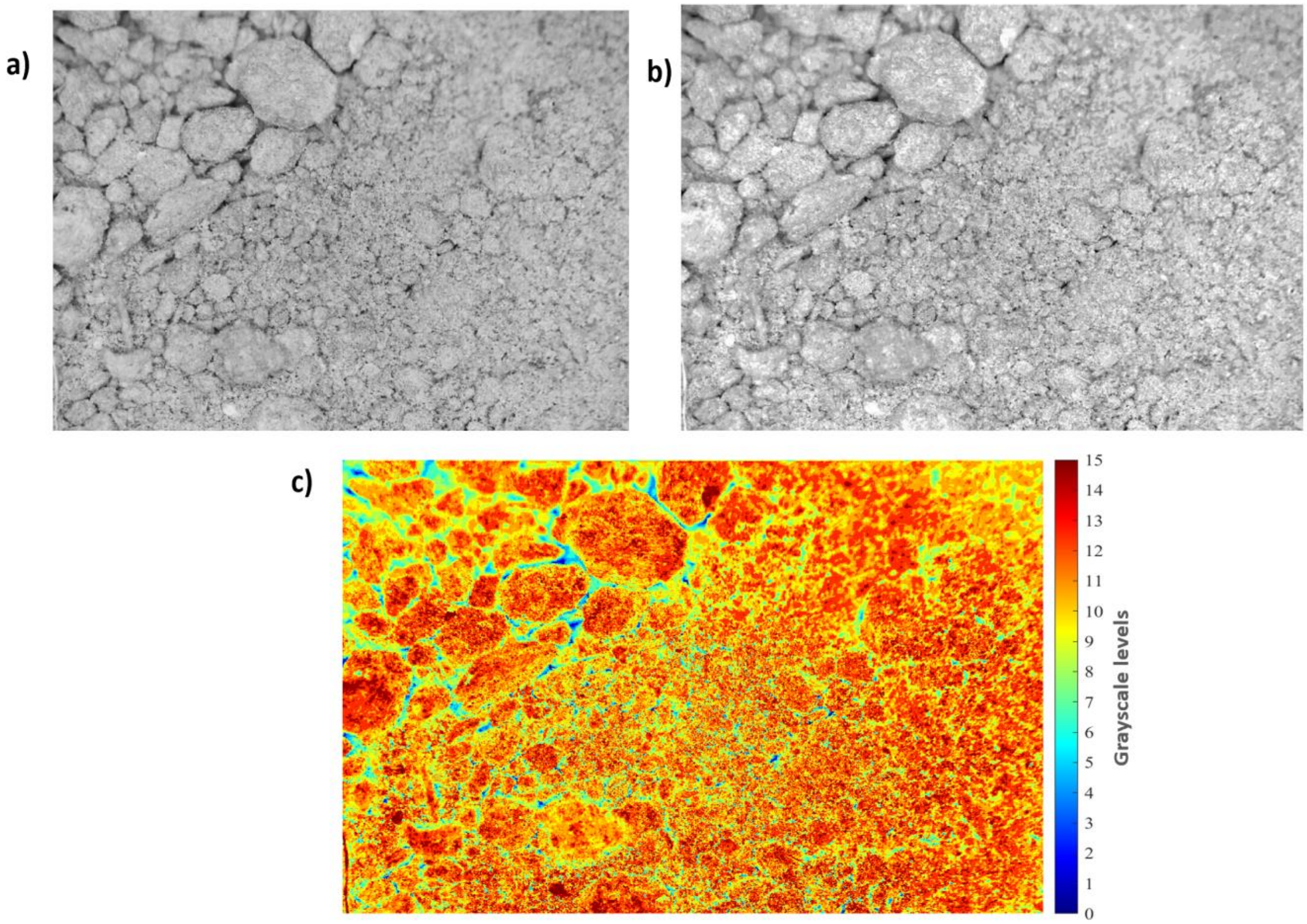

**Figure 3:** Illustration of the image quantization process before computing the gray level co-occurrence matrix (GLCM) and gray level run-length matrix (GLRLM): a) original greyscale (individual channel) image of a soil microscope image, b) image after quantizing the pixel values into 16 levels, and c) the 16 grayscale levels of the image represented on a color map.

### *2.6. Statistical analysis*

The soil image data (n = 1,133) was divided into calibration (80%) and test (20%) sub-datasets. The performance of USB-microscope-based IFs alone, PXRF alone, IFs in conjunction with AVs (IFs + AVs), and IFs and AVs combined with PXRF data (IFs + AVs + PXRF) was assessed for the prediction of OC, available B, available S, available Mn, and SAI. The prediction models employed the random forest (RF) algorithm in Matlab (The Mathworks Inc., Natick, MA), utilizing 300 trees, a minimum leaf size of 8, and a learning rate of 0.1. RF is a ML algorithm that leverages an ensemble of decision trees for predicting continuous values (Breiman, 2001). It builds diverse trees on random subsets of input data, preventing overfitting and enhancing generalization. The algorithm is robust, versatile, and effective for complex regression problems with nonlinear relationships between input and output variables. Its simplicity and minimal hyperparameter tuning make it widely adopted in data science and ML. The RF models were trained with the calibration set (~80%, n = 907 samples) using a 5-fold cross-validation and assessed on the test set (~20%, n = 226 samples). For models predicting B and SAI, a few samples with missing values were excluded. Model performance was assessed using the coefficient of determination ($R^2$), root mean square error (RMSE) (Eq. 1), concordance correlation coefficient ($\rho_c$) (Eq. 2), model bias (Eq. 3), and ratio of performance to inter-quartile distance (RPIQ) (Eq. 4) metrics.

$$RMSE = \sqrt{\frac{1}{n}\sum_{i=1}^{n}(y_i - m_i)^2} \quad (1)$$

$$\rho_c = \frac{2.\rho.\sigma_y.\sigma_m}{\sigma_y{}^2+\sigma_m{}^2+(\mu_y-\mu_m)^2} \quad (2)$$

$$Bias = \frac{1}{n}\sum_{i=1}^{n}(y_i - m_i) \quad (3)$$

$$RPIQ = \frac{Q_3-Q_1}{RMSE} \quad (4)$$

where, $y_i$ and $m_i$ are the actual and predicted values for the $i$-th observation, n is the total number of samples, $\rho$ is the Pearson correlation coefficient between $y$ and $m$, $\sigma_y$ and $\sigma_m$ are the standard deviations of $y$ and $m$, respectively, $\mu_y$ and $\mu_m$ are the means of y and $m$, respectively, and $Q_1$ and $Q_3$ are the first and third quartile of the observed values, respectively.

Principal Component Analysis (PCA) stands as a key dimensionality reduction technique in data analysis and ML (Jolliffe, 2014). Its objective is to transform a dataset into a new coordinate system, where data points are represented by a condensed set of uncorrelated variables known as principal components (PCs). These PCs, sorted by decreasing variance, are linear combinations of the original features. The first PC captures the highest variance, followed by the second PC, and so on. By retaining solely the top few PCs, typically capturing the majority of the variance, PCA enables dimensionality reduction while preserving crucial information. This simplifies data visualization and interpretation and also addresses issues related to high-dimensional data, multicollinearity, and computational complexity. In this study, PCA was conducted to cluster six ACZs using nine IFs (R, G, B, H, S, V, L*, a*, and b*) and combining soil fertility parameters and AVs (OC, available S, available B, SAI, Mn, parent material, and soil order) separately through the 'prcomp' function in R 4.1.2 (R Development Core Team, 2022). Additionally, two more categorical variables (parent material and soil order) were included in PCA. The Filmer-Pritchett procedure, a straightforward approach that breaks down categories into 0/1 dummy variables (Filmer and Pritchett, 2001), was used. This procedure, imposing no additional assumptions on the data, utilized four dummy variables for parent material and three for soil order.

## 3. RESULTS AND DISCUSSION

### *3.1. Descriptive statistics for tested soil parameters*

Table 3 summarizes the soil fertility properties across different ACZs, including available B, OC, available Mn, available S, and SAI. The wide variability in these properties indicates the influence of various factors like parent material, vegetation, geology, topography, soil biota, climate, and land use management in each ACZ (Chakraborty et al., 2019; Swetha et al., 2022).

**Table 3:** Descriptive statistics for laboratory measured soil parameters for soils collected from six different agro-climatic zones of West Bengal, India.

| Statistic | B (mg kg$^{-1}$) | OC (%) | Mn (mg kg$^{-1}$) | S (mg kg$^{-1}$) | SAI |
|---|---|---|---|---|---|
| **Northern Hilly Zone (n=82)** | | | | | |
| Minimum | 1.27 | 1.09 | 18.22 | 4.77 | 5.53 |
| Maximum | 2.11 | 2.85 | 77.19 | 36.39 | 25.94 |
| Mean | 1.71 | 2.06 | 28.75 | 16.18 | 12.17 |
| Std. Deviation | 0.25 | 0.45 | 12.01 | 8.53 | 4.78 |
| Skewness | -0.31 | -0.23 | 2.88 | 0.46 | 1.42 |
| Kurtosis | -1.25 | -0.60 | 10.39 | -0.56 | 2.18 |
| C.V (%) | 15 | 22 | 42 | 53 | 39 |
| **Teesta-Terai Alluvial Zone (n=102)** | | | | | |
| Minimum | 0.73 | 1.02 | 12.57 | 6.03 | 4.32 |
| Maximum | 4.21 | 3.33 | 37.79 | 37.31 | 20.42 |
| Mean | 1.74 | 1.97 | 24.51 | 17.50 | 11.42 |
| Std. Deviation | 0.76 | 0.51 | 6.37 | 7.69 | 3.51 |
| Skewness | 1.53 | 0.13 | 0.30 | 0.38 | 0.41 |
| Kurtosis | 2.69 | -0.35 | -0.91 | -0.62 | -0.01 |
| C.V (%) | 43.78 | 25.87 | 25.97 | 43.96 | 30.77 |
| **Vindhyachal Alluvial Zone (n=214)** | | | | | |
| Minimum | 0.16 | 0.24 | 6.45 | 3.27 | 3.90 |
| Maximum | 1.17 | 1.67 | 155.74 | 83.52 | 43.54 |
| Mean | 0.59 | 0.80 | 46.12 | 16.40 | 11.75 |
| Std. Deviation | 0.27 | 0.32 | 27.00 | 14.64 | 6.61 |
| Skewness | 0.30 | 0.48 | 1.64 | 2.50 | 2.66 |
| Kurtosis | -0.83 | -0.52 | 4.25 | 7.65 | 8.97 |
| C.V (%) | 45.72 | 40.66 | 58.54 | 89.28 | 56.22 |
| **Gangetic Alluvial Zone (n=238)** | | | | | |
| Minimum | 0.23 | 0.22 | 1.18 | 7.29 | 3.29 |
| Maximum | 1.94 | 4.94 | 128.30 | 133.26 | 61.76 |
| Mean | 0.74 | 0.89 | 48.17 | 34.67 | 19.06 |
| Std. Deviation | 0.36 | 0.76 | 26.78 | 27.64 | 11.45 |
| Skewness | 1.17 | 3.92 | 0.32 | 1.83 | 1.61 |
| Kurtosis | 1.30 | 16.61 | -0.54 | 3.41 | 2.86 |
| C.V (%) | 50.00 | 85.51 | 55.60 | 79.72 | 60.07 |
| **Red and Laterite Zone (n=287)** | | | | | |
| Minimum | 0.04 | 0.15 | 3.95 | 1.07 | 2.10 |
| Maximum | 1.78 | 1.83 | 172.46 | 82.80 | 43.11 |
| Mean | 0.74 | 0.64 | 53.74 | 13.12 | 9.77 |
| Std. Deviation | 0.39 | 0.29 | 31.46 | 12.52 | 5.61 |
| Skewness | 0.51 | 1.07 | 1.13 | 2.65 | 2.28 |
| Kurtosis | -0.23 | 2.48 | 1.62 | 9.42 | 8.51 |
| C.V (%) | 52.68 | 45.38 | 58.55 | 95.41 | 57.39 |
| **Coastal Saline Zone (n=210)** | | | | | |
| Minimum | 0.23 | 0.46 | 9.61 | 5.41 | 2.96 |
| Maximum | 3.06 | 1.58 | 139.00 | 350.16 | 145.66 |
| Mean | 0.98 | 0.85 | 45.78 | 39.83 | 20.90 |
| Std. Deviation | 0.68 | 0.24 | 26.05 | 55.16 | 22.28 |
| Skewness | 0.97 | 0.60 | 0.93 | 3.39 | 3.33 |
| Kurtosis | 0.32 | 0.26 | 1.04 | 13.64 | 13.41 |
| C.V (%) | 69.32 | 28.80 | 56.90 | 138.47 | 106.58 |

Among the laboratory-measured properties, available S in the NHZ showed a high coefficient of variation (CV > 50%). The VHZ exhibited high CV values for three properties, and CSZ and RLZ showed high CV values for four properties. Conversely, all five properties in the GAZ displayed high CV values. According to Pimentel-Gomes and Garcia (2002), CV values for available S and SAI indicated significant variability (>30%) across all ACZs, while available Mn and available B also demonstrated considerable variability in five ACZs. Soil OC content ranged widely (0.15 to 4.94%) across ACZs due to variations in soil, climate, and parent materials. The CV of OC varied substantially from 22% to 85.51% across ACZs. Most OC values were below 0.90%, confirming the inherent OC deficiency in Indian soils. The VAZ and RLZ exhibited low mean OC values attributed to their nutrient-poor and gravelly nature. In contrast, NHZ and TAZ displayed very high OC levels. Available B content followed the order TAZ > NHZ > CSZ > GAZ = RLZ > VAZ. NHZ and TAZ exhibited very high B availability (>1.00 mg $kg^{-1}$), attributed to high soil OC content. Three ACZs showed low mean B availability, indicating widespread B deficiency caused by high-yielding varieties, high cropping intensity, inadequate B fertilization, and gradual depletion of native soil mineral sources of B (Saha et al., 2018). Moist sub-humid GAZ and CSZ exhibited high plant-available S content, with GAZ showing a higher range and mean. SAI values varied widely across ACZs, with CSZ having the highest mean SAI value. RLZ exhibited the lowest mean SAI due to leaching loss and low soil organic matter content. For available Mn content, RLZ showed the highest mean value, attributed to metal toxicities and nutrient imbalances. The mean available Mn content in all zones exceeded critical threshold levels. Tropical soils globally reported excess available Mn levels and lower values for available Zn, available B, OC, and cation exchange capacity.

Parent material-wise variation in soil fertility properties was observed (Fig. 4). Different parent materials significantly influenced microbial activity, decomposition rates, and vegetation cover, impacting the distribution and content of SOM. NHZ, with peninsular colluvium, showed higher mean soil OC content, while RLZ, with Granite-Gneiss parent material, showed lower mean OC. Deltaic alluvium parent material exhibited elevated average levels of available B and available S. The correlation plot between PXRF-reported elements and soil fertility parameters highlighted the influence of soil parent material and climate on soil properties and elemental distribution (Fig. 5).

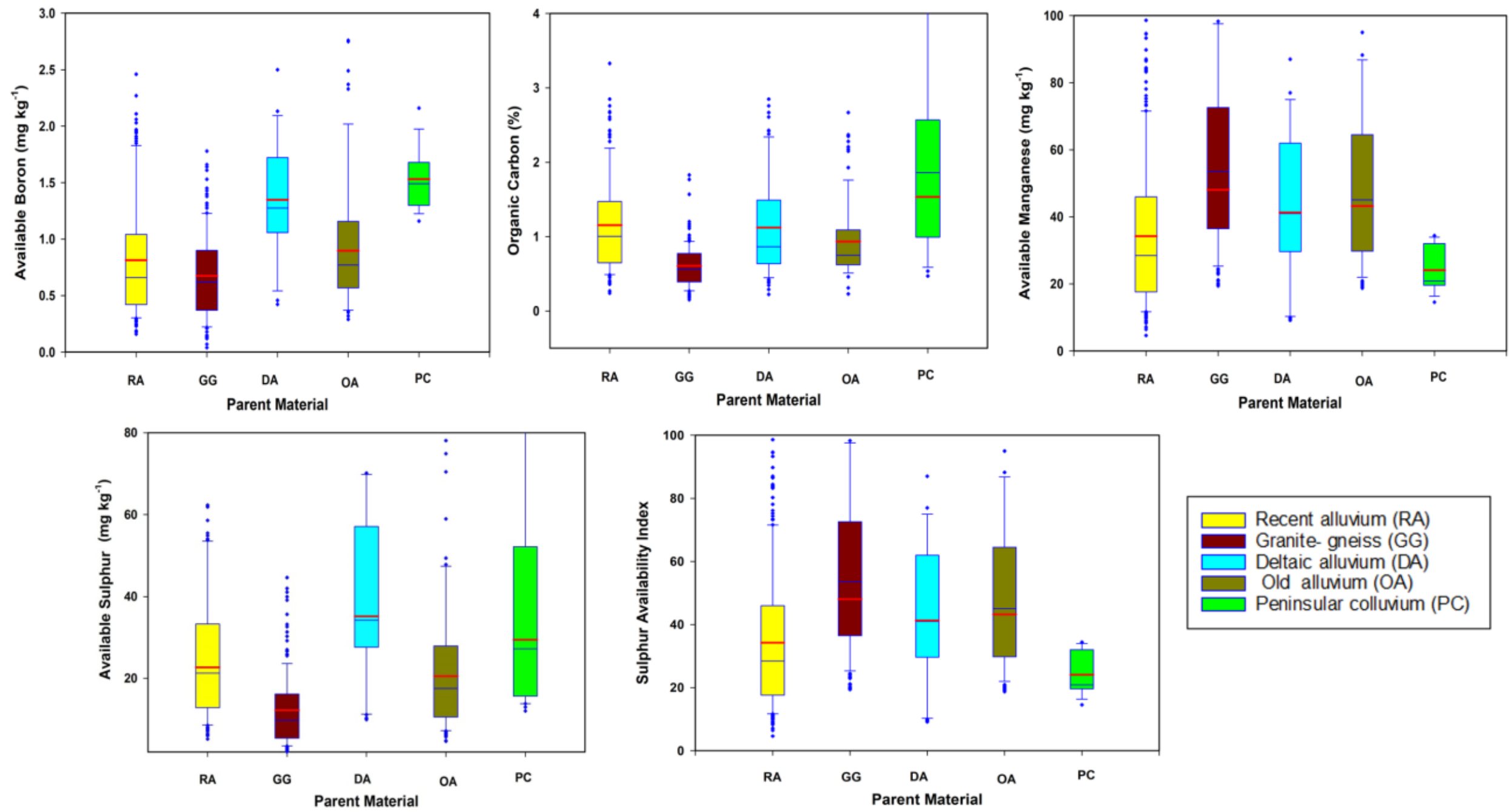

**Figure 4:** Box plots showing distribution of laboratory measured soil parameters among soils representing five soil parent materials of the study region.

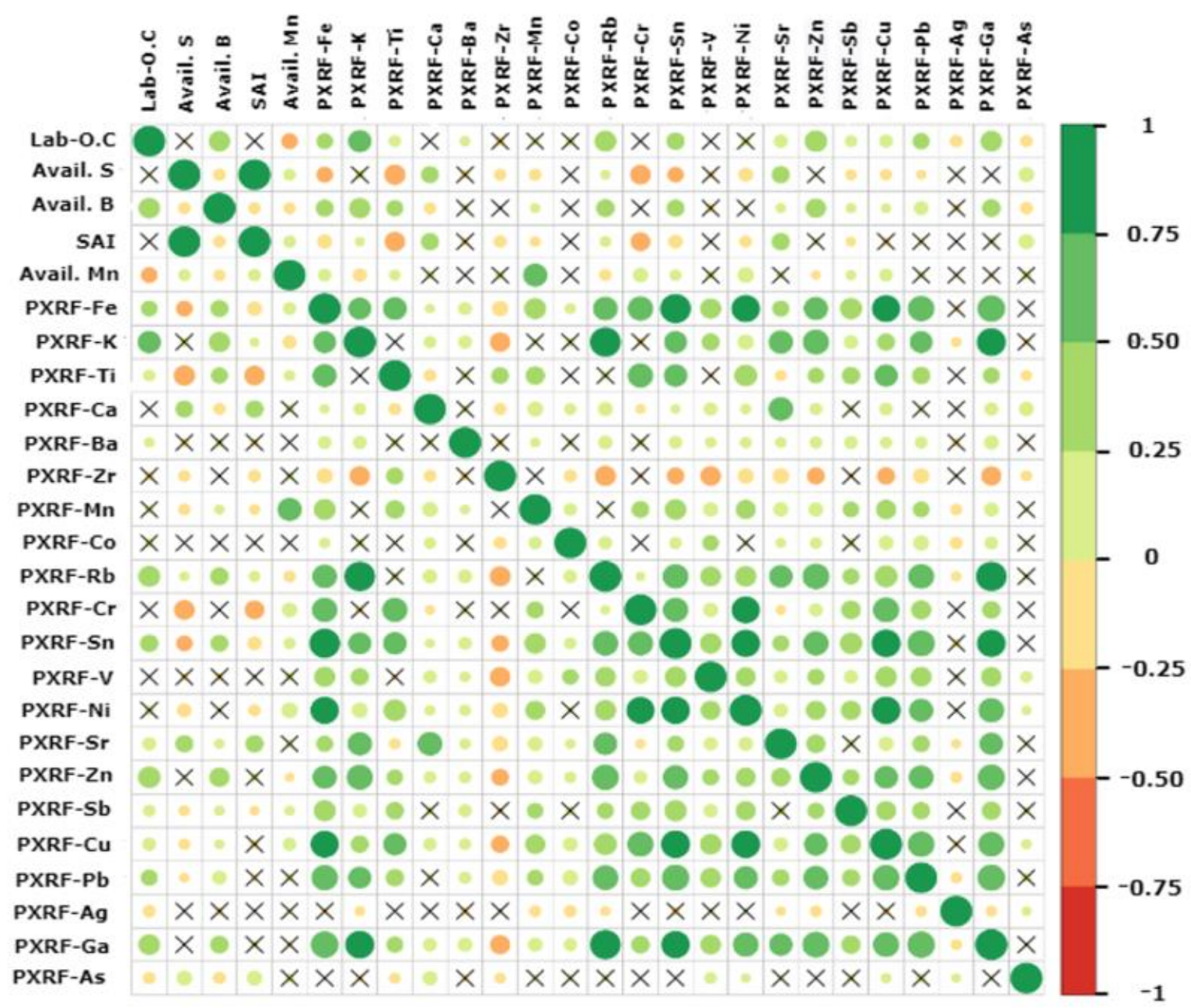

**Figure 5:** Correlations between laboratory measured soil fertility parameters and PXRF elemetal data for soils of West Bengal, India.

### *3.2 Principal component analysis results*

The PCA showed variability in clustering of six ACZs based on IFs alone and a combination of soil fertility and AVs. When using IFs only, the first two PCs explained over 90% of the total variations, with PC1 dominated by the difference between B/b*/a* and S/R/V, and PC2 influenced by the size for G and L (Fig. 6a). Data points clustered in the upper right region, with sub-clusters for samples from different ACZs, such as TTZ and NHZ having larger PC1 values, and RLZ having the lowest PC1 values. The combined PCA (Fig. 6b) showed more scattered samples, with the first two PCs explaining almost 50% of the total variation. PC1 was influenced by S/SAI, highly correlated with acute angles in the biplot, and available B and OC showed good correlation, justifying the trend in Fig. 5. RLZ samples were dominated by Granite-Gneiss and Alfisols. While IFs alone effectively differentiated between ACZs, combining soil fertility variables and AVs resulted in greater variability, forming more diverse clusters, highlighting the influence of these additional variables on sample distribution within ACZs.

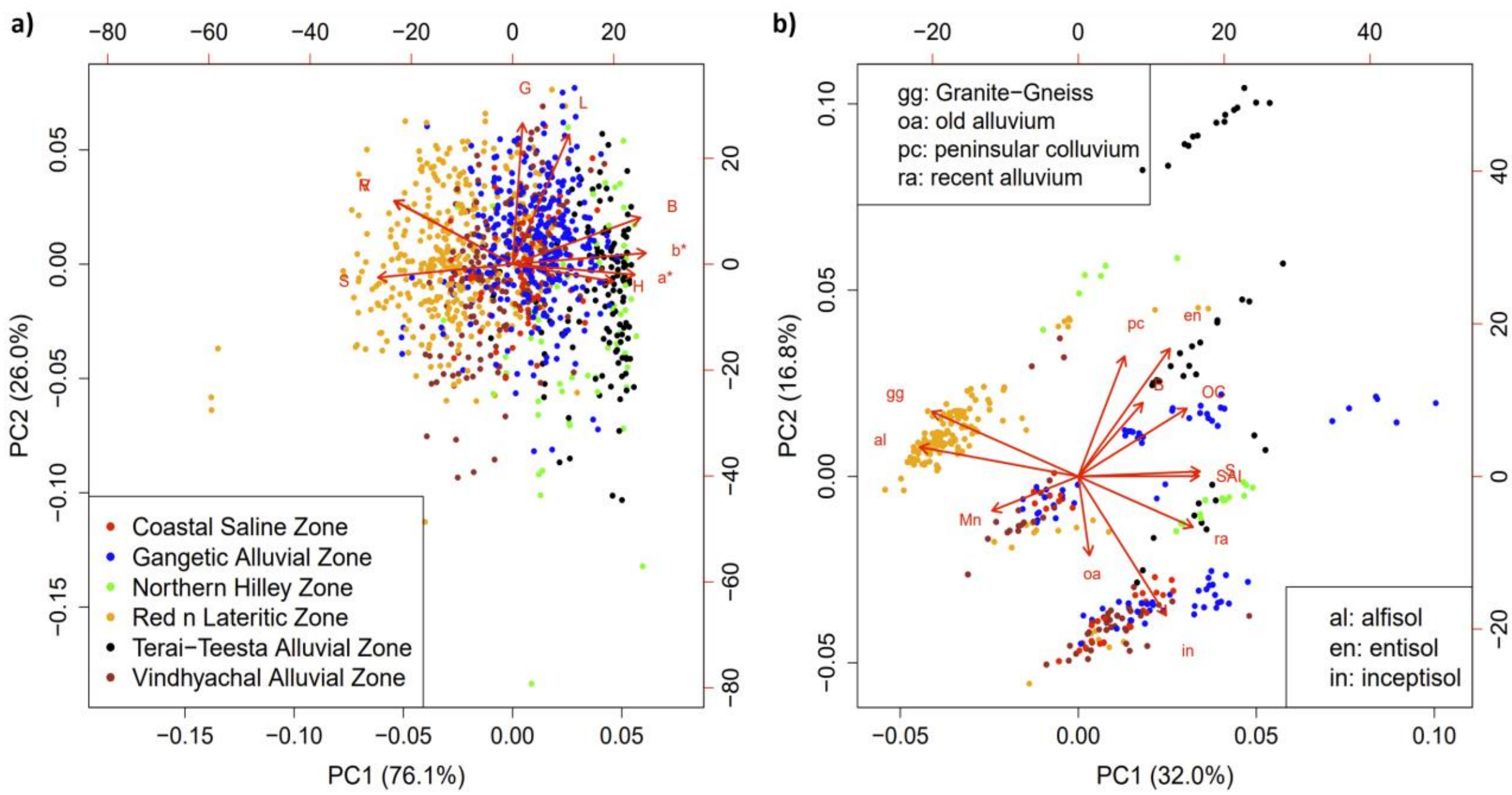

**Figure 6:** The principal component analysis (PCA) biplots using a) microscopic soil image extracted features and agro-climatic zones and b) using laboratory measured soil fertility parameters with soil parent materials and soil orders. Here, S, B and Mn denote available S, available B, and available Mn, respectively.

### *3.3. Prediction model performance for soil fertility parameters*

Initially, the study employed IFs to predict available B, OC, available Mn, available S, and SAI, resulting in test $R^2$ values of 0.36, 0.28, 0.40, 0.33, and 0.41, respectively (Table 4). These findings align with prior research, supporting the assessment of soil attributes based on color or image properties (Aitkenhead et al., 2016 a,b; Swetha et al., 2020; Ymeti et al., 2017). Notably, incorporating AVs as predictor variables significantly enhanced the model's test set performance, particularly for predicting available B (test $R^2$=0.80) and OC (test $R^2$=0.88).

Combining IFs with AVs improved model performances for most fertility parameters compared to using IFs alone, reducing the test RMSE by approximately 60%, 32%, and 17% for OC, available B, and SAI, respectively. However, the prediction performance for available S and available Mn did not show significant changes. These results underscore the impact of the region's agro-climatic conditions and parent material on soil properties, especially for OC. Notably, while prior studies suggested accurate OC prediction using digital IFs alone (Swetha et al., 2020), this study reveals lower accuracy for soil IFs alone, possibly due to the narrow range of targeted soil fertility parameters, limiting the effectiveness of soil color-based prediction (Aitkenhead et al., 2016a). In summary, the study emphasizes the importance of incorporating AVs to establish reliable models applicable over a large geographical area for soil fertility prediction. Conversely, the inclusion of AVs did not affect the prediction of available Mn, indicating the model's ability to predict Mn content without specific knowledge of parent material and ACZ.

**Table 4:** Statistical analysis of the performance of the Random Forest algorithm on different sets of predictor variables via coefficient of determination ($R^2$), root mean square error (RMSE), and bias.

| Response variable | Predictor variables | $R^2$ | | RMSE | | Bias | RPIQ |
|---|---|---|---|---|---|---|---|
| | | Calibration | Test | Calibration | Test | Test | Test |
| B ($mgkg^{-1}$) | IFs + AVs + PXRF | 0.80 | 0.82 | 0.295 | 0.274 | -0.008 | 3.147 |
| | IFs + AVs | 0.78 | 0.80 | 0.314 | 0.296 | -0.008 | 2.914 |
| | PXRF | 0.45 | 0.42 | 0.353 | 0.340 | 0.010 | 2.536 |
| | IFs | 0.35 | 0.36 | 0.448 | 0.432 | -0.010 | 1.997 |
| OC (%) | IFs + AVs + PXRF | 0.84 | 0.87 | 0.248 | 0.234 | 0.032 | 3.034 |
| | IFs + AVs | 0.79 | 0.88 | 0.280 | 0.224 | 0.086 | 3.170 |
| | PXRF | 0.50 | 0.46 | 0.390 | 0.380 | 0.054 | 1.868 |
| | IFs | 0.27 | 0.28 | 0.525 | 0.552 | 0.087 | 1.286 |
| Mn ($mgkg^{-1}$) | IFs + AVs + PXRF | 0.70 | 0.72 | 14.715 | 13.010 | -0.018 | 2.231 |
| | IFs + AVs | 0.48 | 0.44 | 19.248 | 18.333 | -0.383 | 1.582 |
| | PXRF | 0.46 | 0.42 | 19.067 | 18.090 | 0.974 | 1.603 |
| | IFs | 0.47 | 0.40 | 19.000 | 18.891 | 0.905 | 1.535 |
| S ($mgkg^{-1}$) | IFs + AVs + PXRF | 0.54 | 0.50 | 8.268 | 8.453 | 1.358 | 2.162 |
| | IFs + AVs | 0.36 | 0.31 | 9.990 | 10.725 | 2.131 | 1.704 |
| | PXRF | 0.34 | 0.16 | 16.37 | 16.08 | 3.162 | 1.116 |
| | IFs | 0.32 | 0.33 | 10.303 | 10.554 | 2.144 | 1.731 |
| SAI | IFs + AVs + PXRF | 0.67 | 0.70 | 2.985 | 2.855 | -0.190 | 1.951 |
| | IFs + AVs | 0.52 | 0.57 | 3.481 | 3.173 | 0.268 | 1.755 |
| | PXRF | 0.29 | 0.21 | 7.30 | 7.12 | 0.521 | 0.782 |
| | IFs | 0.40 | 0.41 | 3.868 | 3.861 | 0.259 | 1.443 |

IF – Image Feature; AV – Auxiliary Variables

Incorporating PXRF elemental data with (IFs + AVs) notably improved predictions for available S, SAI, and available Mn (Table 4). A substantial enhancement was observed in predicting available Mn, increasing the test $R^2$ from 0.44 to 0.72, and decreasing the test RMSE from 19.24 to 14.71 mg $kg^{-1}$. Similarly, the test RMSE values for available S and SAI decreased by approximately 21% and 11%, respectively. However, the addition of PXRF variables did not improve the prediction accuracy for available B and OC. Thus, while available B and OC can be optimally predicted using (IFs + AVs), combining IFs and AVs with PXRF proved to be more effective for predicting available Mn. Notably, the best results for predicting SAI were achieved with complete sensor fusion (IFs + AVs + PXRF) (test $R^2$ = 0.70; test RMSE = 2.985 mg $kg^{-1}$). Fig. 7 illustrates the predicted vs. measured plots for the target variables using (IFs + AVs + PXRF), where OC indicated a near perfect agreement between observed and predicted values (test $\boldsymbol{\rho_c}$ = 0.93), while a substantial agreement was observed for available B (test $\boldsymbol{\rho_c}$ = 0.88) and available Mn (test $\boldsymbol{\rho_c}$ = 0.81). The other two tested soil fertility indicators exhibited a moderate agreement: available S (test $\boldsymbol{\rho_c}$ = 0.79), and SAI (test $\boldsymbol{\rho_c}$ = 0.75). Nevertheless, the model's performance degrades when predicting higher contents of available Mn (>40 mg $kg^{-1}$), available S (>30 mg $kg^{-1}$), and SAI (>15), possibly due to the lower frequency of soil samples with higher values. Tandon (2011) revealed that the relative scarcity of soils with high levels of available S and SAI values in India can be attributed to low soil organic matter, along with continuous removal of soil S due to intensive cultivation, high production rates, and losses through leaching. Notably, researchers have already showed the S deficiency in RLZ soils (Patra et al., 2012; Sahu et al., 2022). Conversely, samples from the western patch of VAZ exhibited high available Mn; however, their numbers were limited due to low sampling density (Dasgupta et al., 2023). Yet, the tested approach holds promise in characterizing soil fertility for farming communities, enabling the identification of whether their soil fertility is relatively high or low and detecting potential nutritional deficiencies.

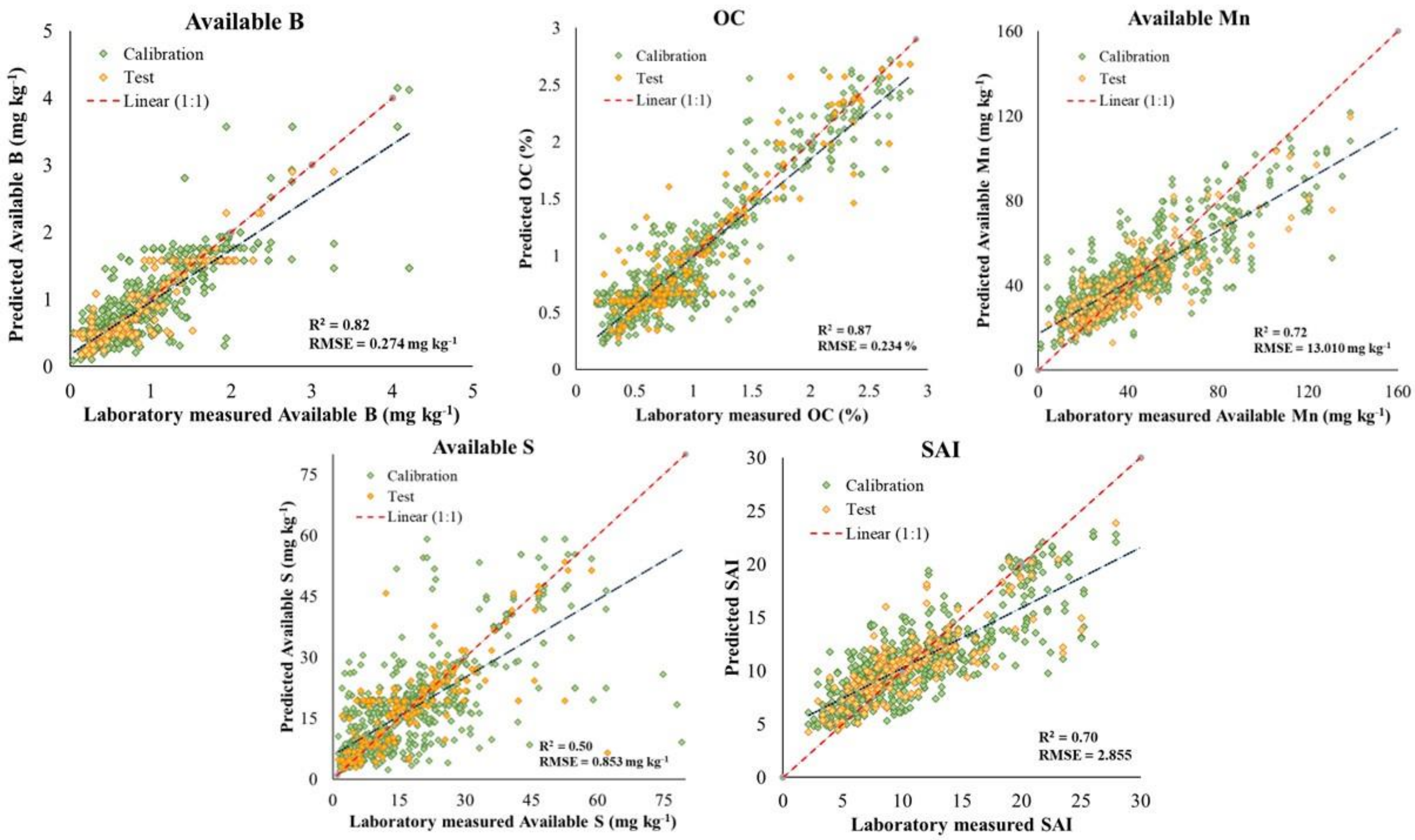

**Figure 7:** Random forest predicted vs laboratory measured plots for the five soil fertility parameters using combined image features, auxiliary variables, and PXRF elemental data. The green and yellow points denote the calibration and test set observations, respectively, while the dotted black and red lines represent the regression line and the 1:1 line, respectively.

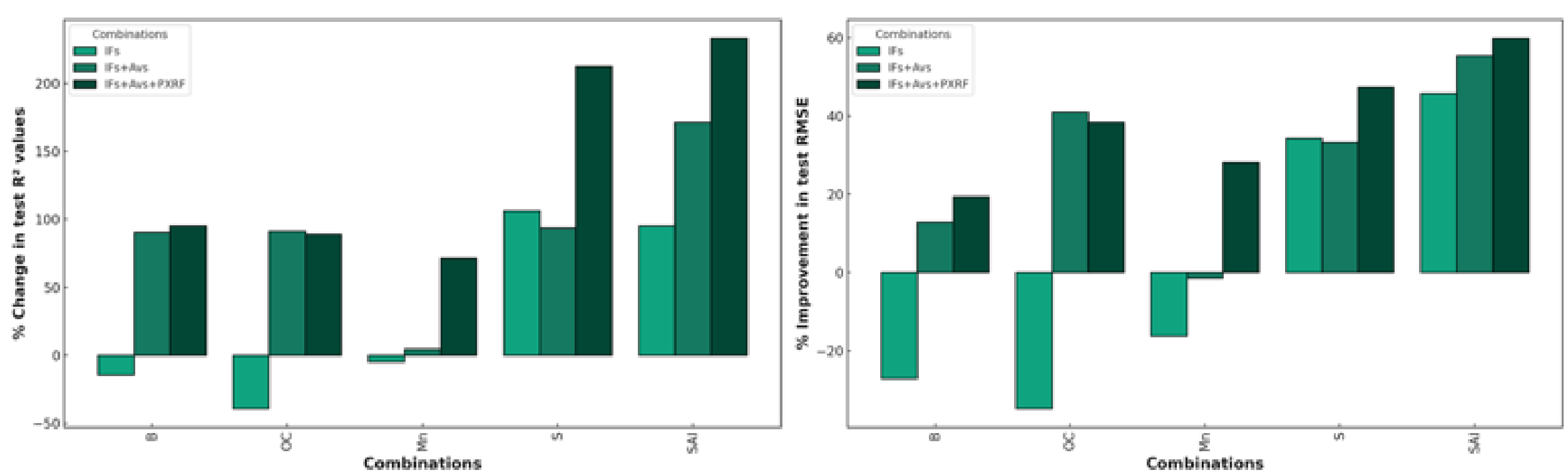

**Figure 8:** Plots showing relative change (%) in soil fertility parameter model a) test $R^2$ and b) test RMSE with microscopic image-extracted features (IFs), auxiliary variables (AVs), and PXRF data in different combinations than using PXRF data alone for soils of West Bengal, India.

[Figure 8](#) showcases the significant improvements in model performance metrics, such as test $R^2$ and test RMSE, achieved by combining image-based sensor data with a sole PXRF-based model. Notably, using IFs alone resulted in a remarkable increase in test $R^2$ values for available S and SAI models (106.25% and 95.24%, respectively) and a simultaneous decrease in model test RMSE (34.37% and 45.77%, respectively) compared to using PXRF data alone. Despite PXRF's direct detection of total Mn, the model that used (IFs + AVs) outperformed the PXRF-based model in predicting available Mn, showing a test $R^2$ improvement of > 4.76%. Interestingly, the addition of a few AVs to IFs significantly enhanced predictive power, with test $R^2$ improvements ranging from 90.48% to 171.43% for available B, OC, available S, and SAI compared to PXRF data alone. However, the dataset encompassing IFs + AVs + PXRF produced high relative improvements in test $R^2$ and reduced test RMSE compared to PXRF or IFs alone for four out of the five tested soil fertility parameters, particularly for available S ($R^2$>212.50%), SAI ($R^2$>233.33%), and available Mn ($R^2$>71.44%).

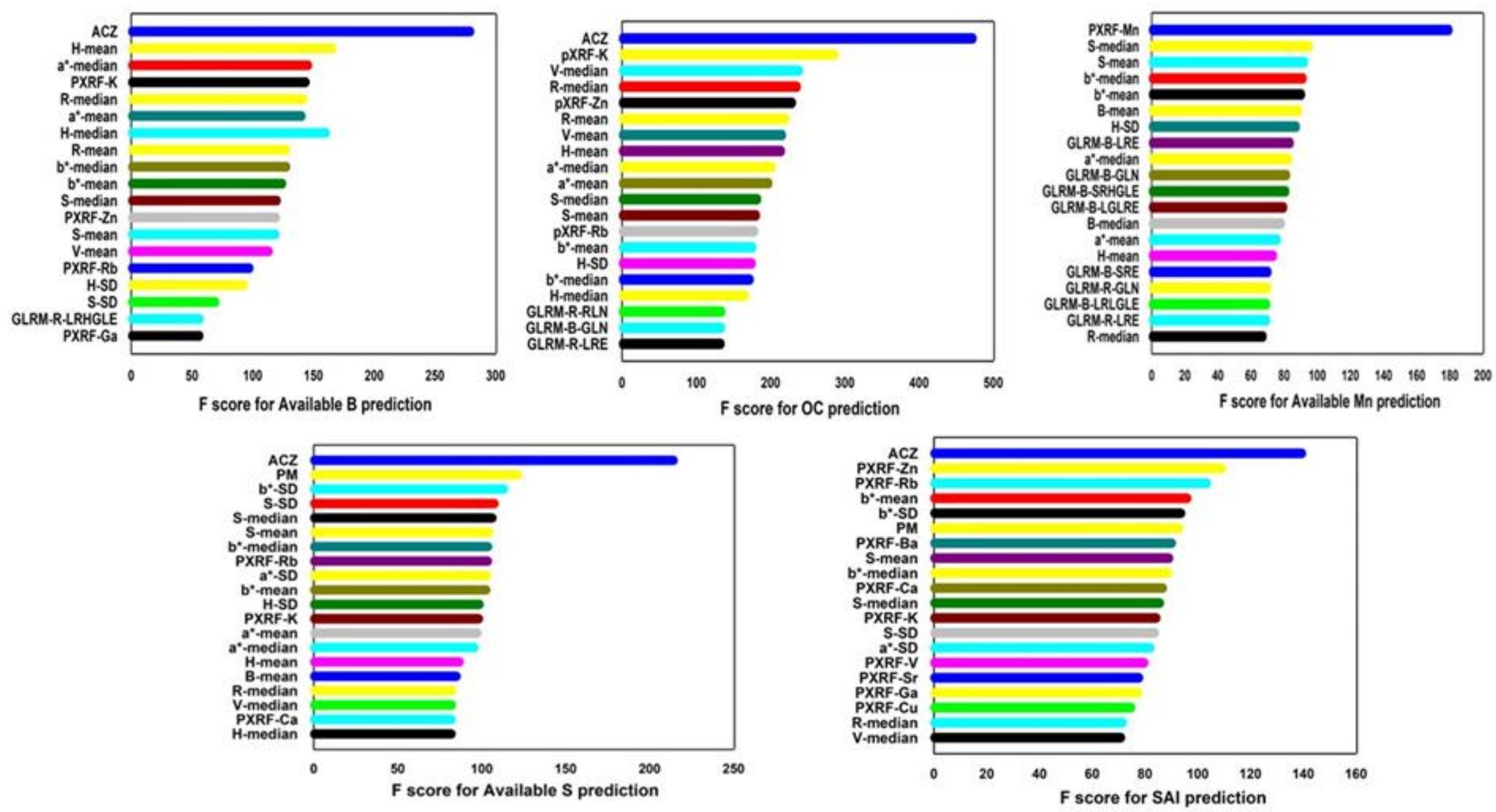

**Figure 9:** Random forest variable importance plots for the prediction of soil fertility parameters. The names of the color and texture feature predictor variables on the y-axis indicate the channel from which the features were derived (R, G, B, H, S, V, L*, a*, b*), followed by the corresponding statistical measure (refer to Table 2). Additionally, texture features are further identified by the matrix from which they were derived (GLCM or GLRM). For example, "GLRM-R-RLN" denotes the run length non-uniformity (RLN) measure derived from the red (R) channel's GLRM.

The feature importance analysis conducted with RF on the complete feature set used for estimating soil fertility properties revealed insightful results ([Fig. 9](#)). The ACZ emerged as the most critical feature for predicting available B, OC, available S, and SAI, indicating its significant role in enhancing prediction performance. Indeed, climatic parameters like rainfall and temperature affects numerous geochemical processes within a certain pH range, altering the solubility of metals from soluble forms (such as $Mn^{2+}$) to insoluble forms (for example, Mn hydroxides) (Brennan et al., 1993; Sauvé et al., 2000; Moreno-Jiménez et al., 2019).

Additionally, color features like H (H – mean, median), a* (a* – mean, median), R (R – mean, median), saturation (S – mean, median) proved dominant for all tested soil properties, emphasizing their importance in the prediction model. The HSV color space and L*a*b* contributed crucial features, and the red channel of RGB (R – mean, median) played a significant role. Specifically, available B prediction relied on features derived from the HSV color space, while OC prediction saw dominance from color features of RGB images. The variable importance plots also highlighted the influence of the R band over G and B, especially for available B and OC. For available S and SAI prediction, b* (b*-mean, median) and saturation (S-mean, median, SD) were more dominant than other color and textural predictors, indicating their close association with sulphate and soil color descriptors. This innovative approach opens new possibilities for predicting soil S-supplying capacity using soil color data within the predictive model, offering valuable insights for rapid soil testing kits, particularly for assessing secondary and micronutrient availability.

This study consistently favored the HSV color model over the RGB model in image processing and predictive modeling, aligning with the findings of Persson (2005) under varying moisture conditions. Key color features for predicting available S and SAI primarily originated from b*, a*, and saturation channels. Textural features derived from GLRLM influenced predictive performance for available Mn but had limited impact on other tested soil nutrients. None of the GLCM features ranked among the top 20 important features, indicating the dominance of color features over image texture features across all tested soil fertility properties. This emphasizes the strong correlation between color features and SOM and mineral composition.

The study highlighted the advantages of incorporating AVs with microscopic image data for soil analysis, supporting previous findings by Dasgupta et al. (2022) using PXRF. Although PXRF cannot analyze OC and light elements, it indirectly provided quantitative information about total elemental compositions, aiding predictions of other soil fertility parameters. The influence of PXRF data was particularly evident for available Mn, available S, and SAI. PXRF-Mn exhibited the highest importance in predicting available Mn, emphasizing its significance. The study revealed the influential role of PXRF-derived elements, such as PXRF-K and PXRF-Rb, in predicting available S, available B, and SAI. PXRF-Zn emerged as a crucial predictor for OC and SAI. The results demonstrated the potential of PXRF and soil imaging systems to provide valuable data for soil fertility prediction.

The enhanced predictive performance of tested soil properties through the incorporation of PXRF data was illustrated in Fig. 10. The correlation between residuals of RF predictions using (IFs + AVs) and RF-predicted values from PXRF data alone indicated improved prediction results, especially for available Mn. ACZ-wise holdout testing revealed differences in prediction accuracies across ACZs, emphasizing the importance of careful dataset partitioning to ensure robust predictions (Table 5). The combination of (IFs + AVs + PXRF) did not significantly enhance prediction accuracy in ACZ-wise holdout test. These findings underscored the need to consider soil type and ACZ variability when partitioning datasets for developing accurate predictive models.

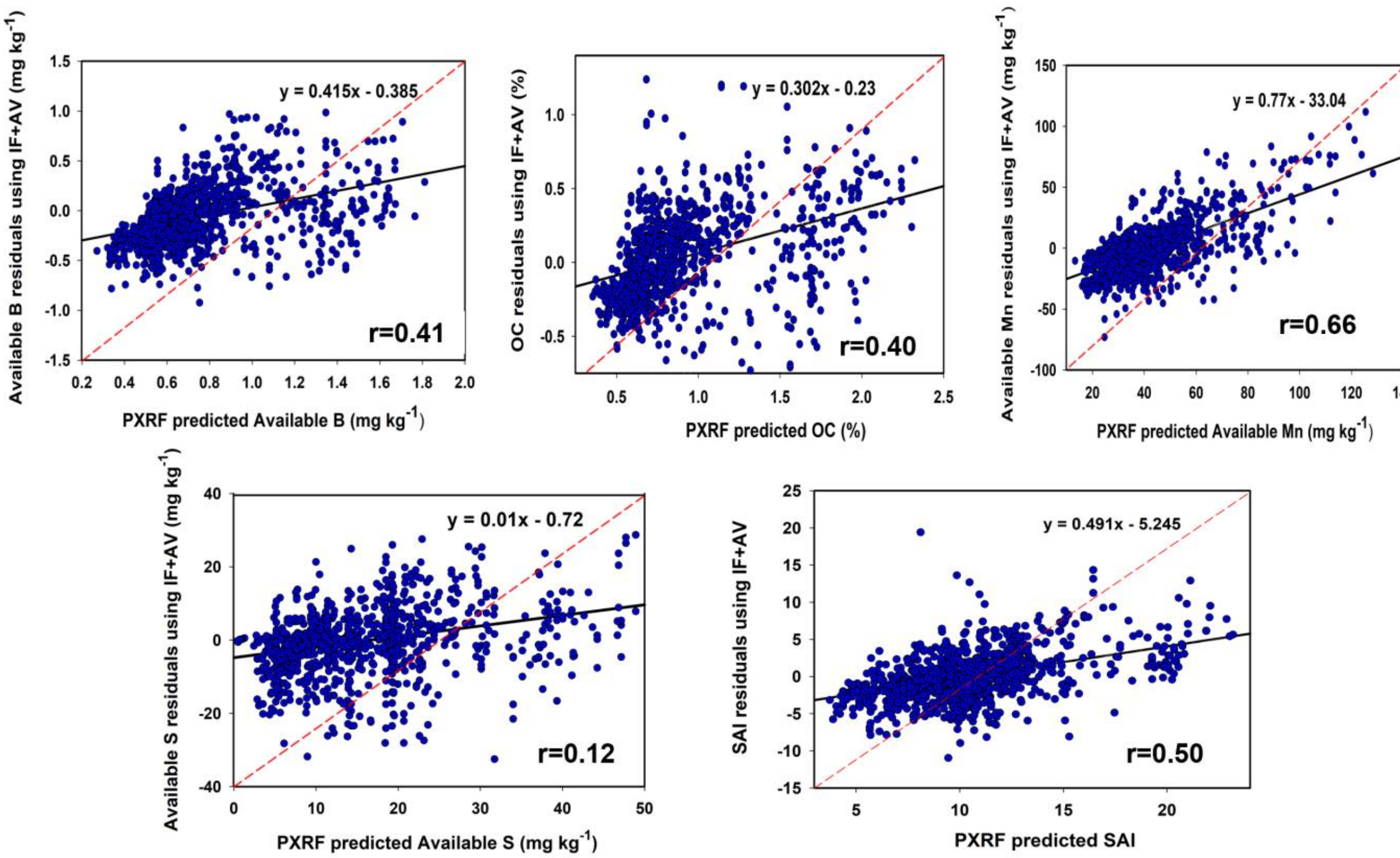

**Figure 10:** The correlation between random forest model reported residuals using IFs + AVs as predictors and predicted soil fertility parameter values using PXRF data alone for soils of West Bengal, India.

**Table 5**: Root mean squared error values for zone-wise holdout test for the prediction of different soil parameters using the Random Forest algorithm for soils from West Bengal, India.

| District | B (mgkg$^{-1}$) | OC (%) | Mn (mgkg$^{-1}$) | S (mgkg$^{-1}$) | SAI |
|---|---|---|---|---|---|
| VAZ | 0.49 | 0.27 | 16.95 | 14.17 | 3.22 |
| NHZ | 1.28 | 1.40 | 22.71 | 20.26 | 6.83 |
| RLZ | 0.43 | 0.34 | 16.74 | 12.56 | 3.02 |
| CSZ | 0.45 | 0.41 | 18.46 | 13.00 | 3.87 |
| GAZ | 0.74 | 0.42 | 20.53 | 17.94 | 4.75 |
| TAZ | 0.92 | 0.59 | 21.10 | 19.16 | 5.28 |

### *3.4. Future scope and challenges of image-based soil fertility prediction*

Digital image processing of color and textural indicators from soil images holds promise for identifying soil geochemistry and mineralogical processes, establishing connections with the agricultural environment. When combined with conventional soil survey data, this soil imagery profile, along with proximal sensor technologies like PXRF or DRS, presents a cost-effective approach to precision agriculture. Integrating physiographic information from soil surveys with in-situ soil microscopic images and using GPS-enabled standardized IDs can facilitate seamless data integration for sensor-based prediction models. These models contribute to soil and environmental studies, offering rapid and reliable data integration for decision support mechanisms related to soil fertility and fertilizer applications. The findings can be incorporated into open-source software development frameworks or IoT platforms, enhancing the long-term utility of the data through server-side image analysis and modeling.

Admittedly, the present study evaluated dried and sieved samples. Hence, to advance this modeling approach, the acquisition of a larger dataset of soil images and scanning soils from a broader range of ACZs in field-moist condition becomes imperative for direct field application. Utilizing advanced image analysis techniques can provide deeper insights and improve prediction accuracy. Integrating soil surface images with remote sensing data and climatic covariates can offer a clearer understanding of how these factors influence soil nutrients and mineralogy on a larger scale, guiding soil nutrient management and policy-making. In further in-situ proximal or imagery studies, efforts should focus on minimizing spatial and temporal variations in soil moisture, extending advanced image processing techniques and optimal model selection to field-moist soil samples. Addressing issues such as heat generation by LEDs leading to condensation on the outer lens in portable microscopes is crucial, and exploring hardware design improvements to mitigate overheating while maintaining constant LED light lux levels is recommended.

## 4. CONCLUSIONS

This study explored the potential of USB microscope cameras as cost-effective and portable tools for detailed soil analysis. Their affordability and magnification capabilities make them valuable for examining soil fractions and color, providing insights into soil micro and secondary nutrient content. The research emphasizes the significance of USB microscope camera-derived IFs as effective predictors for soil properties such as available B, OC, available Mn, available S, and SAI, particularly influenced by soil color and micro-morphological features. The integration of IFs, AVs, and PXRF elemental results underscores the synergies among these variables. The combined use of these predictors enhances the predictive power of models for soil micro and secondary nutrients. Notably, AVs play a crucial role in linking soil properties to environmental factors, contributing to more comprehensive model development. The proposed approach contributes to the advancement of precision agriculture by providing practical and accessible methods for soil analysis and nutrient prediction.

## ACKNOWLEDGEMENTS

The authors acknowledge the All India Coordinated Research Project on Soil Test Crop Response (AICRP on STCR), Indian Council of Agricultural Research (ICAR), Govt. of India for financial assistance.

## 5. REFERENCES

Aitkenhead, M., Coull, M., Gwatkin, R., Donnelly, D., 2016a. Automated soil physical parameter assessment using smartphone and digital camera imagery. J. Imaging 2016, 2 (4), 35. https://doi.org/10.3390/JIMAGING2040035

Aitkenhead, M., Donnelly, D., Coull, M., Gwatkin, R., 2016b. Estimating soil properties with a mobile phone. Digital soil morphometrics 89–110. https://doi.org/10.1007/978-3-319-28295-4_7

Aldabaa, A.A.A., Weindorf, D.C., Chakraborty, S., Sharma, A., Li, B., 2015. Combination of proximal and remote sensing methods for rapid soil salinity quantification. Geoderma 239–240, 34–46. https://doi.org/10.1016/j.geoderma.2014.09.011

Andrade, R., Mancini, M., Teixeira, A.F. dos S., Silva, S.H.G., Weindorf, D.C., Chakraborty, S., Guilherme, L.R.G., Curi, N., 2022. Proximal sensor data fusion and auxiliary information for tropical soil property prediction: Soil texture. Geoderma 422, 115936. https://doi.org/10.1016/j.geoderma.2022.115936

Aygun, S., Gunes, E.O., 2017. A benchmarking: Feature extraction and classification of agricultural textures using LBP, GLCM, RBO, Neural Networks, k-NN, and random forest. 2017 6th International Conference on Agro-Geoinformatics, Agro-Geoinformatics 2017. https://doi.org/10.1109/AGRO-GEOINFORMATICS.2017.8047000

Behera, S.K., Singh, D., Dwivedi, B.S., 2009. Changes in fractions of iron, manganese, copper, and zinc in soil under continuous cropping for more than three decades. Commun. Soil Sci. Plant Anal. 40, 1380–1407. https://doi.org/10.1080/00103620902818054

Berger, K.C., Truog, E., 1940. Boron deficiencies as revealed by plant and soil tests1. Agron. J. 32, 297–301. https://doi.org/10.2134/agronj1940.00021962003200040007X

Breiman, L., 2001. Random forests. Mach. Learn. 45, 5–32. https://doi.org/10.1023/A:1010933404324

Brennan, R.F., Armour, J.D., Reuter, D.J., 1993. Diagnosis of zinc deficiency. In: Robson, A.D. (Ed.), Zinc in soils and plants Springer, Dordrecht, pp. 167-181.

Chakraborty, S., Li, B., Weindorf, D.C., Deb, S., Acree, A., De, P., Panda, P., 2019. Use of portable X-ray fluorescence spectrometry for classifying soils from different land use land cover systems in India. Geoderma 338, 5–13. https://doi.org/10.1016/j.geoderma.2018.11.043

Chakraborty, S., Weindorf, D.C., Michaelson, G.A.R.Y.J., Ping, C.L., Choudhury, A., Kandakji,

T., Acree, A., Sharma, A., Wang, D., 2016. In-situ differentiation of acidic and non-acidic tundra via portable x-ray fluorescence (PXRF) spectrometry. Pedosphere 26, 549–560. https://doi.org/10.1016/S1002-0160(15)60064-9

Choudhury, B.U., Divyanth, L.G., Chakraborty, S., 2023. Land use/land cover classification using hyperspectral soil reflectance features in the Eastern Himalayas, India. Catena, 229, 107200. https://doi.org/10.1016/J.CATENA.2023.107200

Dasgupta, S., Chakraborty, S., Weindorf, D.C., Li, B., Silva, S.H.G., Bhattacharyya, K., 2022. Influence of auxiliary soil variables to improve PXRF-based soil fertility evaluation in India. Geoderma Reg. 30, e00557. https://doi.org/10.1016/j.geodrs.2022.E00557

Dasgupta, S., Debnath, S., Das, A., Biswas, A., Weindorf, D.C., Li, B., Kumar Shukla, A., Das, S., Saha, S., Chakraborty, S., 2023. Developing regional soil micronutrient management strategies through ensemble learning based digital soil mapping. Geoderma 433, 116457. https://doi.org/10.1016/j.geoderma.2023.116457

Dasgupta, S., Sengupta, S., Saha, S., Sarkar, A., Anantha, K.C., 2021. Approaches in advanced soil elemental extractability: catapulting future soil-plant nutrition research. Soil Sci. Fundam. to Recent Adv. 191–236. https://doi.org/10.1007/978-981-16-0917-6_10/COVER

Divyanth, L.G., Rathore, D., Senthilkumar, P., Patidar, P., Zhang, X., Karkee, M., Machavaram, R., Soni, P., 2023. Estimating depth from RGB images using deep-learning for robotic applications in apple orchards. Smart Agricultural Technology 6, 100345. https://doi.org/10.1016/J.ATECH.2023.100345

Donahue, R.L., Miller, R.W., Shickluna, J.C., 1977. Soils : an introduction to soils and plant growth. Prentice-Hall, Englewood, Cliffs, N.J.

Duda, B.M., Weindorf, D.C., Chakraborty, S., Li, B., Man, T., Paulette, L., Deb, S., 2017. Soil characterization across catenas via advanced proximal sensors. Geoderma 298, 78–91. https://doi.org/10.1016/j.geoderma.2017.03.017

Filmer, D., Pritchett, L.H., 2001. Estimating wealth effects without expenditure data - Or tears: An application to educational enrollments in states of India. Demography 38, 115–132. https://doi.org/10.2307/3088292

Fu, Y., Taneja, P., Lin, S., Ji, W., Adamchuk, V., Daggupati, P., Biswas, A., 2020. Predicting soil organic matter from cellular phone images under varying soil moisture. Geoderma 361, 114020. https://doi.org/10.1016/j.geoderma.2019.114020

Gorthi, S., Swetha, R.K., Chakraborty, S., Li, B., Weindorf, D.C., Dutta, S., Banerjee, H., Das, K., Majumdar, K., 2021. Soil organic matter prediction using smartphone-captured digital images: Use of reflectance image and image perturbation. Biosyst. Eng. 209, 154–169. https://doi.org/10.1016/j.biosystemseng.2021.06.018

Goswami, R., Dutta, S., Misra, S., Dasgupta, S., Chakraborty, S., Mallick, K., Sinha, A., Singh,

V.K., Oberthür, T., Cook, S., Majumdar, K., 2023. Whither digital agriculture in India? Crop and Pasture Science 74, 586-596. https://doi.org/10.1071/CP21624

Jolliffe, I., 2014. Principal component analysis. In: Balakrishnan, N., Colton, T., Everitt, B., Piegorsch, W., Ruggeri, F., Jozef, L., Teugels (Eds.), Wiley StatsRef: Statistics Reference Online, John Wiley & Sons, pp. 2-29.

Lindsay, W.L., Norvell, W.A., 1978. Development of a DTPA soil test for zinc, iron, manganese, and copper. Soil Sci. Soc. Am. J. 42, 421–428. https://doi.org/10.2136/SSSAJ1978.03615995004200030009X

Majumdar, K., Ray, D.P., Chakraborty, S., Pandit, T., 2014. Change of nutrient status of hilly soil in Darjeeling district within five years. Int. J. Bioresour. Sci. 1, 25–30.

Mathew, A., Antony, A., Mahadeshwar, Y., Khan, T., Kulkarni, A., 2022. Plant disease detection using GLCM feature extractor and voting classification approach. Mater Today Proc 58, 407–415. https://doi.org/10.1016/J.MATPR.2022.02.350

Minasny, B., McBratney, A.B., 2016. Digital soil mapping: A brief history and some lessons. Geoderma 264, 301–311. https://doi.org/10.1016/J.GEODERMA.2015.07.017

Moinuddin, G., Jash, S., Sarkar, A., Dasgupta, S., 2017. Response of potato (Solanum tuberosum L.) to foliar application of macro and micronutrients in the red and lateritic zone of West Bengal. J. Crop Weed 13, 185–188.

Moreno-Jiménez, E., Plaza, C., Saiz, H., Manzano, R., Flagmeier, M., Maestre, F.T., 2019. Aridity and reduced soil micronutrient availability in global drylands. Nature Sustainability 2(5), 371-377.

Mouazen, A.M., Karoui, R., Deckers, J., De Baerdemaeker, J., Ramon, H., 2007. Potential of visible and near-infrared spectroscopy to derive colour groups utilising the Munsell soil colour charts. Biosyst. Eng. 97, 131–143. https://doi.org/10.1016/J.BIOSYSTEMSENG.2007.03.023

Nadimi, M., Divyanth, L.G., Paliwal, J., 2023. Automated Detection of Mechanical Damage in Flaxseeds Using Radiographic Imaging and Machine Learning. Food Bioproc Tech 16, 526–536. https://doi.org/10.1007/S11947-022-02939-5/FIGURES/9

Nayak, D.C., Sarkar, D., Velayutham, M., 2001. Soil Series of West Bengal. Publication No. 89. National Bureau of Soil Survey and Land Use Planning, Nagpur.

Padhan, D., Sen, A., Rout, P.P., 2016. Extractability and availability index of sulphur in selected soils of Odisha. J. Appl. Nat. Sci. 8, 1981–1986. https://doi.org/10.31018/JANS.V8I4.1074

Patra, P., Mondal, S., Ghosh, G.K., 2012. Status of available sulphur in surface and sub-surface soils of red and lateritic soils of West Bengal. International Journal of Plant, Animal and Environmental Sciences 2(2), 276–281.

Persson, M., 2005. Estimating surface soil moisture from soil color using image analysis. Vadose

Zo. J. 4, 1119–1122. https://doi.org/10.2136/VZJ2005.0023

Pimentel-Gomes, F., Garcia, C.H., 2002. Estatística aplicada a experimentos agronômicos e florestais: exposição com exemplos e orientações para uso de aplicativos 309.

Qi, L., Adamchuk, V., Huang, H.H., Leclerc, M., Jiang, Y., Biswas, A., 2019. Proximal sensing of soil particle sizes using a microscope-based sensor and bag of visual words model. Geoderma 351, 144–152. https://doi.org/10.1016/J.GEODERMA.2019.05.020

R Development Core Team, 2022. R: A language and environmental for statistical computing. R Found. Stat. Comput.Available online at: http://www.Rproject.org (Verified 29 June 2022).

Rowe, W. F., 2005. Forensic applications. In: Hillel, D. (Ed.), Encyclopedia of Soils in the Environment, Elsevier, Amsterdam, pp. 67-72. https://doi.org/10.1016/B0-12-348530-4/00495-1

Saha, A., Mani, P.K., Hazra, G.C., Dey, A. and Dasgupta, S., 2018 . Determining critical limit of boron in soil for wheat (Triticum aestivum L.). Journal of plant nutrition 41(16), 2091-2102

Saha, S., Saha, B., Seth, T., Dasgupta, S., Ray, M., Pal, B., Pati, S., Mukhopadhyay, S.K., Hazra, G., 2019. Micronutrients availability in soil–plant system in response to long-term integrated nutrient management under rice–wheat cropping system. J. Soil Sci. Plant Nutr. 19, 712–724. https://doi.org/10.1007/S42729-019-00071-6

Sahu, M., Nisab, M.C.P., Mondal, S., 2022. Availability of macronutrients & sulphur and their relationship with physico-chemical properties in lateritic soils of Birbhum district, West Bengal. International Journal of Economic Plants 9(3), 260-263.

Sarkar, S., Gaydon, D.S., Brahmachari, K., Poulton, P.L., Chaki, A.K., Ray, K., Ghosh, A., Nanda, M.K., Mainuddin, M., 2022. Testing APSIM in a complex saline coastal cropping environment. Environ. Model. Softw. 147. https://doi.org/10.1016/J.ENVSOFT.2021.105239

Sauvé, S., Hendershot, W., Allen, H.E., 2000. Solid-solution partitioning of metals in contaminated soils: dependence on pH, total metal burden, and organic matter. Environmental science and technology 34(7), 1125-1131.

Shukla, A.K., Behera, S.K., Prakash, C., Tripathi, A., Patra, A.K., Dwivedi, B.S., Trivedi, V., Rao, C.S., Chaudhari, S.K., Das, S., Singh, A.K., 2021. Deficiency of phyto-available sulphur, zinc, boron, iron, copper and manganese in soils of India. Sci. Rep. 11, 19760. https://doi.org/10.1038/S41598-021-99040-2

Singh, M. V., 2008. Micronutrient deficiencies in crops and soils in India. Micronutr. Defic. Glob. Crop Prod. 93–125. https://doi.org/10.1007/978-1-4020-6860-7_4/COVER

Sudarsan, B., Ji, W., Adamchuk, V., Biswas, A., 2018. Characterizing soil particle sizes using wavelet analysis of microscope images. Comput. Electron. Agric. 148, 217–225.

https://doi.org/10.1016/J.COMPAG.2018.03.019

Sudarsan, B., Ji, W., Biswas, A., Adamchuk, V., 2016. Microscope-based computer vision to characterize soil texture and soil organic matter. Biosyst. Eng. 152, 41–50. https://doi.org/10.1016/J.BIOSYSTEMSENG.2016.06.006

Swetha, R., Chakraborty, S., Dasgupta, S., Li, B., Weindorf, D.C., Mancini, M., Silva, S.H.G., Ribeiro, B.T., Curi, N., Ray, D.P., 2022. Using nix color sensor and munsell soil color variables to classify contrasting soil types and predict soil organic carbon in eastern India.Comput. Electron. Agric. 199, 107192. https://doi.org/10.1016/j.compag.2022.107192

Swetha, R.K., Bende, P., Singh, K., Gorthi, S., Biswas, A., Li, B., Weindorf, D.C., Chakraborty, S., 2020. Predicting soil texture from smartphone-captured digital images and an application. Geoderma 376, 114562. https://doi.org/10.1016/j.geoderma.2020.114562

Swetha, R.K., Chakraborty, S., 2021. Combination of soil texture with Nix color sensor can improve soil organic carbon prediction. Geoderma 382, 114775. https://doi.org/10.1016/j.geoderma.2020.114775

Tandon, H.L.S., 2011. Sulphur in soils, crops and fertilizers. Fertilizer development and consultation organization (FDCO), New Delhi.

Taneja, P., Vasava, H.K., Daggupati, P., Biswas, A., 2021. Multi-algorithm comparison to predict soil organic matter and soil moisture content from cell phone images. Geoderma 385, 114863. https://doi.org/10.1016/j.geoderma.2020.114863

Viscarra Rossel, R.A., Fouad, Y., Walter, C., 2008. Using a digital camera to measure soil organic carbon and iron contents. Biosyst. Eng. 100, 149–159. https://doi.org/10.1016/j.biosystemseng.2008.02.007

Walkley, A., Black, I.A., 1934. An examination of the Degtjareff method for determining soil organic matter, and a proposed modification of the chromic acid titration method. Soil Science 37, 29–38. https://doi.org/10.1097/00010694-193401000-00003

Wang, D., Chakraborty, S., Weindorf, D.C., Li, B., Sharma, A., Paul, S., Ali, M.N., 2015. Synthesized use of VisNIR DRS and PXRF for soil characterization: Total carbon and total nitrogen. Geoderma 243–244, 157–167. https://doi.org/10.1016/j.geoderma.2014.12.011

Weindorf, D.C., Chakraborty, S., 2020. Portable X-ray fluorescence spectrometry analysis of soils. Soil Sci. Soc. Am. J. 84, 1384–1392. https://doi.org/10.1002/SAJ2.20151

Ymeti, I., Van Der Werff, H., Shrestha, D.P., Jetten, V.G., Lievens, C., van der Meer, F., 2017. Using color, texture and object-based image analysis of multi-temporal camera data to monitor soil aggregate breakdown. Sensors, 17(6), 1241. https://doi.org/10.3390/S17061241

Zenda, T., Liu, S., Dong, A., Duan, H., 2021. Revisiting Sulphur—The once neglected nutrient: it's roles in plant growth, metabolism, stress tolerance and crop production. Agric. 11(7),

626. https://doi.org/10.3390/agriculture11070626